# Electrically controlled localized charge trapping at amorphous fluoropolymer-electrolyte interfaces

Hao Wu [*,a,b,c], Ranabir Dey [c,e], Igor Siretanu [c], Dirk van den Ende [c], Lingling Shui [a, b], Guofu Zhou [a,b,d], Frieder Mugele [*,c]

[a] Guangdong Provincial Key Laboratory of Optical Information Materials and Technology & Institute of Electronic Paper Displays, South China Academy of Advanced Optoelectronics, South China Normal University, Guangzhou 510006, P. R. China.

[b] National Center for International Research on Green Optoelectronics, South China Normal University, Guangzhou 510006, P. R. China.

[c] Physics of Complex Fluids, Faculty of Science and Technology, MESA+ Institute for Nanotechnology, University of Twente, Enschede 7500AE, The Netherlands

[d] Shenzhen Guohua Optoelectronics Tech. Co. Ltd., Shenzhen 518110, P. R. China

[e] Present address: Dynamics of Complex Fluids, Max Planck Institute for Dynamics and Self-organization, Am Fassberg 17, 37077 Goettingen, Germany.

KEYWORDS: Charge trapping, electrowetting, Teflon AF, hydrophobic, controllable charge pattern

ABSTRACT: Charge trapping is a long-standing problem in electrowetting-on-dielectric (EWOD), causing reliability reduction and restricting its practical applications. Although this phenomenon has been investigated macroscopically, the microscopic investigations are still lacking. In this work, the trapped charges are proven to be localized at three-phase contact line region by using three detecting methods -- local contact angle measurements, electrowetting (EW) probe, and Kelvin Probe Force Microscopy (KPFM). Moreover, we demonstrate that this EW-induced charge trapping phenomenon can be utilized as a simple and low-cost method to deposit charges on fluoropolymer surfaces. Charge density near the three-phase contact line up to 0.46 mC/m$^2$ and the line width with deposited charges ranging from 20 to 300 µm are achieved by the proposed method. Particularly, negative charge densities do not degrade even after "harsh" testing with a water droplet on top of the sample surfaces for 12 hours, as well as after being treated by water vapor for 3 hours. These findings provide an approach for applications which desire stable and controllable surface charges.

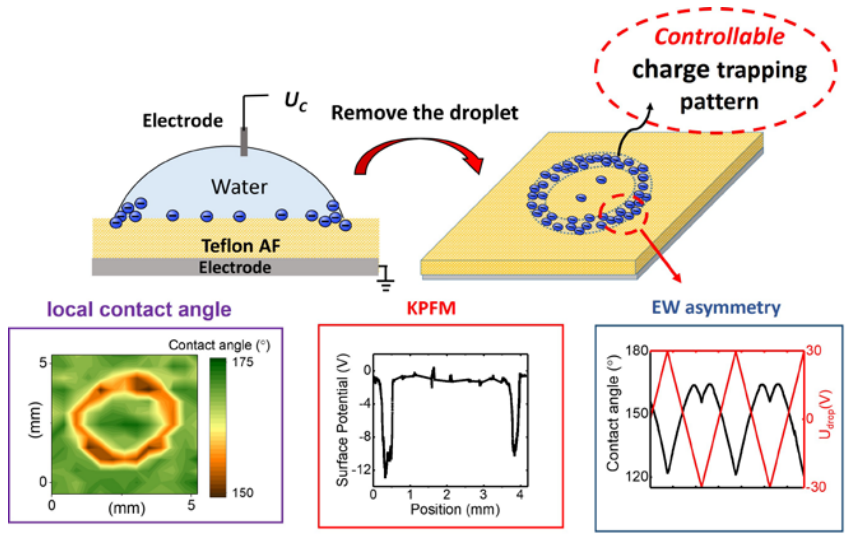

*Table of Contents Graphic*

## 1. INTRODUCTION

Amorphous fluoropolymers (AFPs) such as Teflon AF and Cytop are popular materials for various applications[1-7] because of the unique combination of favorable material properties, such as chemical inertness, mechanical strength, water repellency, dielectric strength, optical transparency, and easy solution processability [7-8]. For these reasons, AFPs are also predominantly used as insulating and hydrophobic layer in electrowetting (EW) devices [8-14]. EW, which is often also denoted as 'electrowetting-on-dielectric' (EWOD) to emphasize the relevance of the dielectric layer, relies on the fact that ionic charge carriers have in general a rather low affinity to the weakly polarizable AFPs. However, at the same time, fluoropolymers have been used for decades as charge storage (electret) materials with applications in electro-mechanical transductions, such as microphones, micro-electro-mechanical systems (MEMS) and electric generators[15-19]. These applications rely on the fact that charges, once deposited on or within AFPs, remain stable due to the wide electronic band gap and deep energetic traps. The purpose of the present work is to shed light on these two antagonistic aspects of charge repellence and charge storage in AFPs that jointly control the injection of charge carriers into AFPs in EWOD at high voltage. Such EW-induced charge injection, if done in a controllable way, will offer interesting opportunities for generating permanent charge patterns on AFPs.

The reliability of any EW applications in microfluidics[20-21], optofluidics[22-23], display technology[24-25], and energy harvesting[26] relies on the reproducibility, performance, and durability of the dielectric layer; thus the stability of AFPs is particularly important[27-30]. Charge trapping induces the degradation of the electrical response of AFP films, leading to contact angle saturation and failures in EWOD devices[26, 31-32]. Early experiments with composite dielectrics displayed a reversible response and symmetric saturation for positive and negative bias voltage, suggesting substantial mobility of both types of charge carriers upon injection into the AFP films[33]. More recent studies displayed a strongly asymmetric and sometimes irreversible response[34,35]. Other investigations identified a relationship of macroscopic charge injection and/or contact angle saturation with molecular scale properties of the system, demonstrating better EW stability for bulkier salt ions[29] (e.g. surfactants, ionic liquids) and for bulk fluid molecules[36] (e.g. glycols). More recently, it has been also reported that Teflon AF materials got permanent negative surface charges upon extended (several hours) contact with water in the absence of any applied voltage[37].

Since most studies on EW-induced charging phenomena mainly focus on the response of the macroscopic contact angles and the total electrical currents, there is a significant lack of quantitative description of the underlying microscopic charge trapping phenomenon. Moreover, a clear understanding of the correlation between the charge trapping and the macroscopic wetting characteristics has remained elusive. It has been recognized that diverging electric fields in the vicinity of the three phase contact line (TPCL) cause various types of non-linear response of the materials during EW, which may limit the minimum contact angle[38-39]. Nevertheless, it is still not clear whether the heterogeneity of the electric fields leads to the charge trapping and induces permanent change in the local surfaces of AFPs. It was assumed that the charge injection process essentially follow the distribution of the electric field with its well-established divergence near the TPCL [40-41]. However, several recent studies have revisited the topic of charge injection in EWOD using local surface potential measurements with non-contact electrostatic probes [32, 42-43]. Surprisingly, the measured surface potential distributions were reported to be rather broad with a maximum in the center of the droplet, thereby challenging the classical view based on the local field divergence[32, 42-43].

In this work, we analyze the charge distribution generated on AFP surfaces by EW at high voltage with unprecedented lateral resolution, and explore the usage of the EW-induced charge injection for localized charge storage at AFP surfaces. Three complementary techniques have been used to reveal the local charge distribution on single layer AFP surfaces. We demonstrate that EW-induced charge injection is highly localized. Based on this, a simple and low-cost approach is proposed and validated to generate stable charge patterns with controllable length scale and density. As a result, we can tune the surface properties of AFP surfaces at microscale level without complex microfabrication processes and the related instruments. The excellent stability of the negative trapping charges, in particular in a "harsh" environment under water or high humidity, suggests its potential for a wide range of applications requiring stable surface charges[44-48].

## 2. EXPERIMENTAL SECTION

2.1 Preparation of Teflon films

ITO/glass substrates were cleaned in a Liquid Crystal Display (LCD) cleaning line for G2.5 glass (400 mm × 500 mm). Subsequently, 800 nm thick AFP films were prepared by screen printing Teflon AF 1600 solution (The Chemours Company, USA), followed by baking on a hot plate at 95 °C for 1 min to remove residual solvent and additional baking in an oven at 185 °C for 30 min to anneal the film. All processes were carried out in a clean room. More details on the fabrication process can be found in Ref.[49].

2.2 Surface charging

Teflon surfaces were charged by applying DC voltages $U_c$ of up to ± 140 V for 2 to 15 min between the ITO electrodes on the substrate (kept at electrical ground potential) and a platinum (Pt) wire (0.1 mm diameter) immersed into a 5 µL drop of de-ionized water (MilliQ). Charging the surface was performed at room temperature (~25 °C) in a closed container filled with vapour-saturated air. The generic setup is shown in Figure S1 along with a time trace showing the suppression of drop evaporation by the water-saturated atmosphere. Electrical voltages were generated by a function waveform generator (33220A, Agilent, USA) in combination with an amplifier (PZD 700, Trek, USA). The charging voltage $U_c$ was typically chosen within the range of contact angle saturation, in which $\theta$ depends only weakly on the applied voltage (see Figure S2). $U_c$ was limited to ensure that the simultaneously measured current on the substrate remained below 1 µA for all experiments.

2.3 Surface characterization

*Contact angle measurements*

The wettability of the samples was measured using a commercial contact angle goniometer (OCA-15+, Data Physics, Germany). Advancing and receding contact angles of the Teflon surfaces in air prior to charging were $\theta_a^{air} = 120°$ and $\theta_r^{air} = 115°$. All subsequent contact angle measurements after charging with and without electrowetting (EW) were carried out in ambient silicone oil (317667, Sigma-Aldrich, USA) with probe droplets of 0.3-0.5 µL, substantially smaller than the charging drop. All reported voltages are measured with respect to the grounded ITO electrodes. The water drop was in contact with the Pt wire at all time; *i.e.* contact angles at zero voltage corresponding to a configuration with an electrically grounded droplet (i.e. the droplet is not freely floating). The advancing and receding contact angles under these conditions were close to 170° with negligible hysteresis. The EW response was probed by applying a triangular waveform ($\pm 30$ V) with a period of 60 s to the probe drop. The maximum voltage during the EW-surface characterization measurements was kept deliberately low to ensure that the system displays a parabolic response following the equation

$$\cos\theta(U) = \cos\theta_Y + \frac{c}{2\gamma}(U - U_T)^2 \qquad (1)$$

Here, $\theta$ is the contact angle under applied voltage of $U$, and $c = \frac{\varepsilon_0 \varepsilon_r}{d} = 2.2 \times 10^{-5}\ F/m^2$ is the capacitance per unit area between the liquid and the electrode on the substrate. $\gamma = 41\ mJ/m^2$ is the oil-water interfacial tension. ($\varepsilon_0 \varepsilon_r$ and $d \approx 800 nm$: dielectric permittivity and thickness of Telfon AF layer). Following our previous work[37], we denote the offset voltage $U_T$ as the 'trapping' voltage. It corresponds to the potential on the droplet when it contains zero charge (*i.e.* all the counter charges are remain in the bottom electrode). The observation of a finite offset voltage directly points to the presence of a finite permanent surface charge density of

$$\sigma_T = c\ U_T \qquad (2)$$

at the polymer-electrolyte interface[33, 37]. This formula implicitly assumes that the deposited charge with density of $\sigma_T$ resides on the top of the Teflon surface and does not penetrate substantially into the bulk of the material. Penetration to a depth $\delta$ would lead to an enhanced capacitance $c(\delta) = c \cdot d/(d - \delta)$. However, as long as $\delta$ is only a few nm, *i.e.* small fraction of $d$, the resulting correction would thus be minor. In addition, a finite charge on the surface automatically implies that $\theta(U = 0) < \theta_Y$. Correspondingly, there is a finite screening charge density $\sigma_D(U = 0) = \sigma_T \frac{c_{EDL}}{c+c_{EDL}} \approx \sigma_T$ on the drop ($c_{EDL} \gg c$: electric double layer capacitance). In contrast, we denote $\theta_Y = \theta(U_T)$ as Young's angle, which corresponds to the contact angle of zero charge on the drop. For more detailed aspects of EW measurements including their interpretation in the presence or absence of surface charges, see Ref. [8].

*Kelvin probe force microscopy (KPFM)*

To characterize the electrostatic potential of the surface in more detail and with high lateral resolution, KPFM measurements were performed using a commercial atomic force microscopy (AFM; Dimension Icon Bruker, USA) with conductive (Sb-(n)doped Si) rectangular tips with a nominal tip radius of 25 nm (SCM-PIT-V2, BRUKER, USA). Upon applying an AC voltage ($U_{AC} = 500\ mV,\ f = 60{\sim}62 kHz,$ ) superimposed onto a DC voltage ($U_{DC}$) to the AFM tip, the electrostatic force ($F_{es}$) between the AFM tip and sample is given by:

$$F_{el} = \frac{1}{2}\frac{\partial C(z)}{\partial z}(U_{DC} - U_T + U_{AC} \sin \omega t)^2 \qquad (3)$$

Here $\partial C(z)/\partial z$ is the gradient of the capacitance between tip and sample surface and $U_T$ is the trapping voltage. Splitting the force according to their frequency ($\omega$), we obtain the static ($F_{DC}$) and dynamic ($F_\omega$ and $F_{2\omega}$) contributions, as usual,

$$F_{DC} = \frac{\partial C(z)}{\partial z}\left[\frac{1}{2}(U_{DC} - U_T)^2 + \frac{1}{4}U_{AC}^2\right] \qquad (4)$$

$$F_\omega = \frac{\partial C(z)}{\partial z}(U_{DC} - U_T)U_{AC} \sin \omega t \qquad (5)$$

$$F_{2\omega} = -\frac{1}{4}\frac{\partial C(z)}{\partial z}U_{AC} \cos 2\omega t \qquad (6)$$

The amplitude of $F_\omega$ is proportional to $U_{DC} - U_T$. To obtain the $U_T$ in amplitude modulation KPFM, $U_{DC}$ is adjusted such that $F_\omega$ becomes minimal. For a system with a perfectly homogeneous dielectric film and a bottom electrode layer, the surface potential $U_S$ is expected to be identical with the trapping voltage $U_T$. More details on measuring $U_T$ with the KPFM can be found in S.I. and Figure S3.

## 3. RESULTS AND DISCUSSION

3.1 Evidence of charge trapping phenomenon

*Macroscopic surface wettability upon charging*

The schematic and working principle of EWOD is shown in Figure 1a. When a voltage ($U$) is applied on the dielectric layer via a electrolyte droplet and the bottom electrode, a pulling force ($f_U$) from the applied electric field pulls the TPCL toward the outside direction of the droplet, and thus changes the contact angle. This pulling force $f_U = \frac{\sigma_s^2}{2c}$ is governed by the total amount of charges at the electrolyte/solid interfaces $\sigma_s$ (the same amount of counter charges are in the bottom electrode). When the applied voltage ($U$) is relatively low and all surface charges are supposed to be contributed by the electric field, the pulling force can be written as: $f_U = \frac{1}{2}cU^2$. This is the explanation of the classical Young-Lippmann EW model. However, when the voltage reaches a certain high value, charges can be injected at the electrolyte/solid interfaces, and thus causing contact angle saturation. The charge trapping process could be reversible or irreversible. If the charges are irreversibly trapped in the dielectric layer, the electric response of the dielectric layer will be permanently degraded.

To investigate the charge trapping phenomenon in EWOD, we deposited a 5 µL drop of de-ionized water on the sample surface in air, as described above, and abruptly turned the charging voltage to as high as $-120\ V$. As a response, the drop spread within a few tens of ms from the initial contact angle of ~115° to ~ 70° (Figure 1b). Subsequently, a slow relaxation to $\theta(U_{ch}) \approx$ 80° took place for approximately 1 min. Along with the increase in contact angle, the radius of footprint area of the drop decreased (Figure 1c). This macroscopic contact angle retreat phenomenon has also been observed in previous reports[28, 34, 42]. According to the humid environment, this relaxation was not caused by evaporation (see Figure S1). After 300 s, when the charging voltage was turned off, subsequent inspection of the samples by optical and by atomic force microscopy (Figure 4c) did not display any appreciable variation of the surface topography.

As discussed above, the contact angle variation in EW is the joint effect of the materials surface (interface) tension, the applied voltage, the reversible and the irreversible trapping charges. In the present experiment, the material surface (interface) tension and the applied voltage ( $U = -120$ V) were kept constant, and the pulling force contributed by the applied voltage of was $f_U = \frac{1}{2}cU^2 = 158$ mN/m. Due to the effect of reversible and irreversible trapping charges, the pulling forces were suppressed to $\gamma_{w/a}(\cos 115° - \cos 70°) = 55$ mN/m for the initial contact angle of 70 ° to $\gamma_{w/a}(\cos 115° - \cos 80°) = 43$ mN/m for the final contact angle of 80 ° (water/air interfacial tension $\gamma_{w/a} = 72$ mN/m). The 12 mN/m reduction of pulling force is supposed to be caused by the charge density variation of both reversible and irreversible charge trapping.

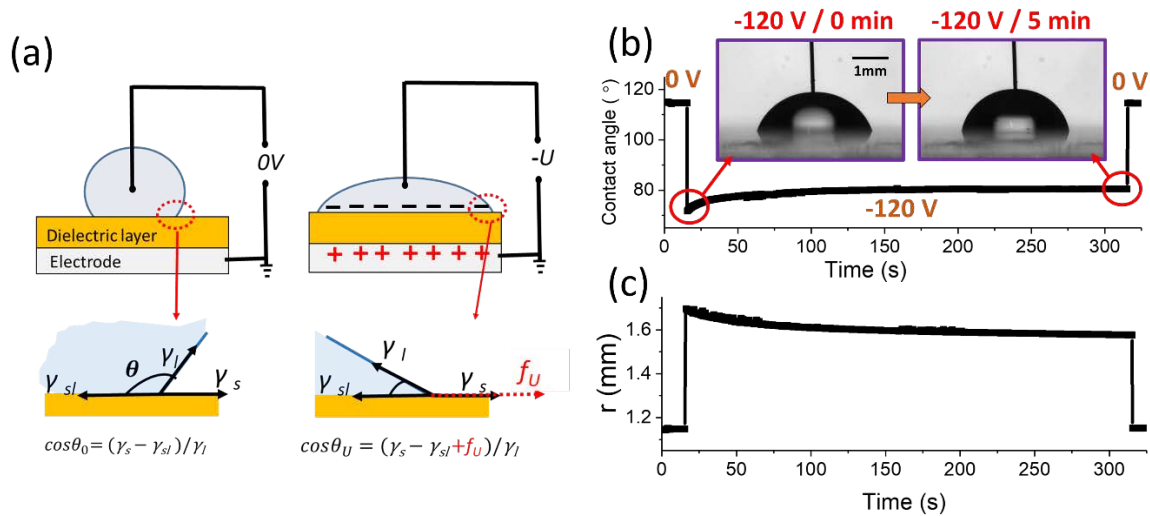

**Figure 1.** (a) Illustration of the electrowetting on dielectric (EWOD) principle. $\gamma_s$, $\gamma_l$ and $\gamma_{sl}$ are the solid/air, liquid/air and solid/liquid interface tensions. (b) Water/air contact angle and (c) contact line radius of a 5 µL water droplet depending on time with -120 V voltage applied. Insets of b) show side view images of charging droplets immediately after applying the voltage and dafter 5 min. This charging process is shown in Video S1.

*Local contact angles*

To investigate whether the charge trapping was reversible and spread at the entire drop-substrate interface, we removed the charging drop (after turning off the charging voltage) and subsequently investigated the surface properties in several manners. First, the wettability of the surface were investigated at high lateral resolution using a contact angle measurement with a much smaller probe droplet (0.3 µL) (the setup shown in Figure S4). To minimize disturbing effects of contact angle hysteresis, these measurements were carried out in ambient oil. Figure 2a shows a composite side view image of such a probe droplet at various locations on the surface. Video S2 shows the complete process of this local contact angle measurement. The probe drop was always in contact with the electrically grounded Pt wire, guaranteeing zero potential drop between the drop and the electrode on the substrate. Away from the position of the charging drop, the contact angle of the probe drop was close to 170°. Interestingly, in the center of the charging drop, the same contact angle was observed. Within the resolution of this measurement, this part of the surface thus behaved the same as the pristine outer parts that were not exposed to the charging drop at all. In contrast, in the region close to the contact line during charging, $\theta(U=0)$ was reduced to about 155°. The width of this region with reduced contact angle was approximately 0.2 to 0.5 mm. Figures 2b shows a one-dimensional variation of theta with position along a diameter of the charging droplet footprint; Figure 2c shows a full two-dimensional map of the reduction of $\theta(U)$ all along the contact line of the original charging drop. The variation of the contact line corresponded to an pulling force changing in oil was obtained to be $\Delta f_o = \gamma(\cos 155° - \cos 170°) = 0.08\,\gamma \approx 3.2 \text{ mN/m}$. Assuming $\Delta f_o$ was contributed by the trapped charges, governing by $\Delta f_o = \sigma_T{}^2/2c$, the surface charge density was around 0.37 mC/m².

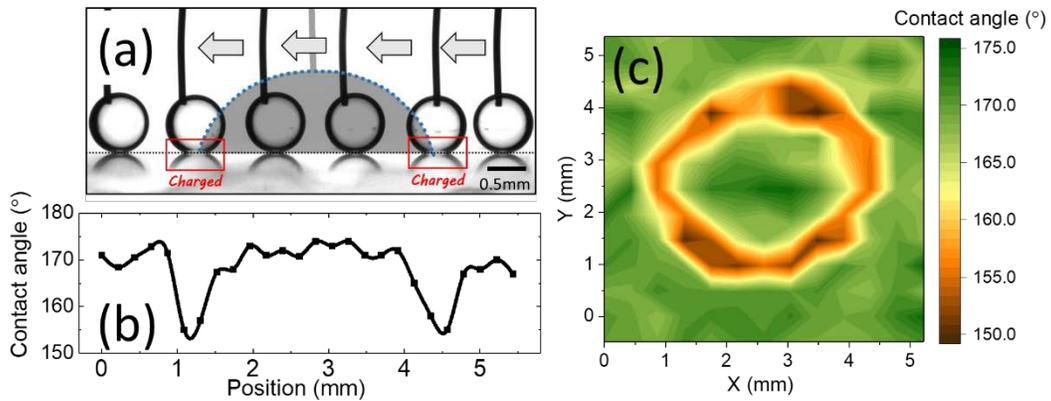

**Figure 2.** Contact angle reduction along contact line of charging drop. (a) Snapshots of electrically grounded probe-droplets ($V = 0.3\ \mu L$) at different locations relative to the original charging droplet (grey background; $V = 5\ \mu L$) that was used to charge the surface ($U_c = -120\ V; t_c = 5\ min$). This test is also shown in Video S2. (b) Contact angle $\theta(U = 0)$ of electrically grounded drop vs. position extracted from (a) . (c) 2D map of $\theta(U = 0)$ (12x13 locations) as measured by probe drop.

*Electrowetting response*

The reduction of $\theta(U = 0)$ presented in Figure 2 alone does not clearly indicate whether the effect was caused by local chemical variation of the surface along the contact line, which would reduce Young's angle $\theta_Y$ in eq. (1), or whether it is indeed caused by the expected injection of surface charge, which would give rise to a finite value of $U_T$. To distinguish between these two scenarios, we performed EW measurements using smaller probe droplets (0.5 µL). Ramping the voltage applied to the droplet up and down in triangluar fashion, we found that the decrease in contact angle with increasing voltage was asymmetric, as expected in the presence of a static surface charge (Figures 3b and c). The asymmetry was found to be strongly position-dependent on te surface, being much more pronounce close to the contact line during charging, Region 2 in Figure 3, as compared to the central part of the charging drop (Region 1). At the same time, maximum contact angle, the apex of the curves, corresponding to the contact angle at zero charge was also slightly decreased as compared to the pristine case. Fitting the eq. (1) to the data shown in Figure 3c, we found that the trapping voltages were of $U_T(1) = -16V$ and $U_T(2) = -3V$ in Regions 1 and 2, respectively. The corresponding trapping charge density could be calculated by Eq. (2), and the $\sigma_T$ were 0.34 and 0.06 mC/m² in Regions 1 and 2. $\sigma_T$ of 0.34 mC/m² at Region 1 was consistent with the calculated values from the θ(U = 0) variation in Figure 2, which was 0.37 mC/m². The consistance of $\sigma_T$ value calculated from local contact angle measurment and the the EW response indicates that the reduction of $\theta(U = 0)$ (Figure 2) is induced by the irrevesible charge trapping, instead of the chemical surface modification. Hence, the AFP surface modification has been achieved via this procedure, with the deposition of charges by a physical way rather than a chemical modification.

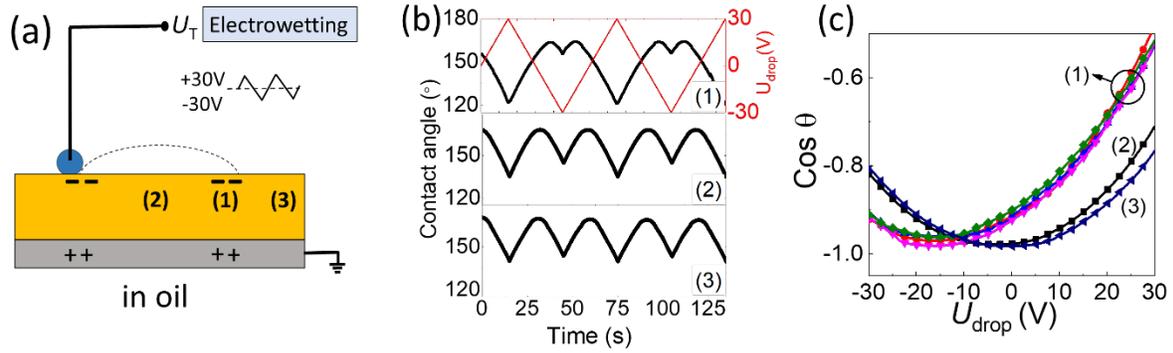

**Figure 3.** (a) Schematic of measuring trapping charge by electrowetting probe with different probing regions (1): contact line; (2) drop center; (3) pristine surface. (b) EW response curves and (c) contact angle vs. applied voltage ($U_{drop}$) all three regions. (charging conditions: -120 V for 5 min).

*Kelvin Probe Force Microscopy (KPFM) measurement*

Considering the fact that the probe droplets spread upon applying the voltage, and thus their footprint area increases quickly, one may wonder whether the probe drop remains within the narrow ring around the original contact line where the deposited charges are presumably trapped. If the probe drop spreads beyond the charging area, the measured asymmetry of the EW response and the value of $U_T$ would in part reflect the finite lateral extent of the deposited charge pattern rather than its absolute value.

To overcome the resolution limitation of contact angle-base detecting method, we performed AFM and KPFM measurements on the prepared charged Teflon surfaces in ambient air after removing the charging drop without immersing the surface into oil. The sample was charged by −90 V for 5 min. The AFM topography images displayed a very smooth surface with a roughness of a few nanometers. No indications of topographic surface modifications due to the charging process could be identified. In contrast, the KPFM measurements indeed revealed strong lateral variations of the surface potential $U_S$ in the region of the contact line, as shown in Figure 4. Overall, the KPFM measurements confirmed the important observations of the macroscopic surface characterization (Figures 2 and 3). $U_S$ was essentially constant and small in the center of the charging drop-substrate interface. Pronounced variations of $U_S$ occurred in the rim along the contact line of the original charging drop. This rim was around 200 µm to 300 µm wide, only slightly smaller than that suggested by the contact angle and EW response measurements. The absolute value of the local surface potential in the KPFM measurements was around -10 V which was consistent with the EW-response measurements at the same charging conditions (shown in Figure S5 and 5a). Nano-sized KPFM probe in ambient air and macroscopic EW-probed drops in ambient oil thus experienced the same surface charge density $\sigma_T$, which could be obtained from the measured voltages using eq. (2) with $U_T = U_S$. Thus far, these results from three types of micro- and nano-scale measurements revealed and confirmed that the charges are indeed trapped at the AFP surfaces after EW process and accumulate at the TPCL regions.

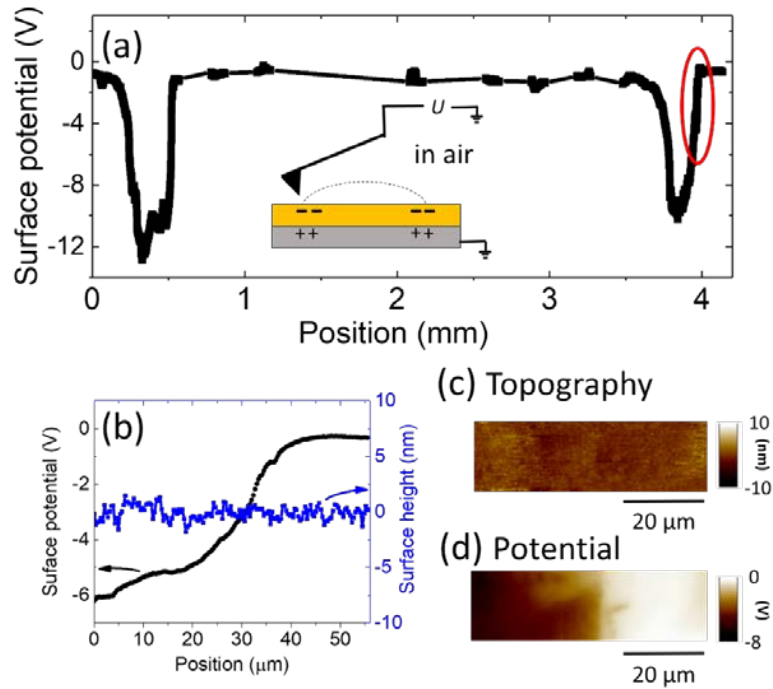

**Figure 4.** (a) Surface potential map measured by stitching several KPFM measurements of $90 \times 90$ $\mu m^2$. (charging conditions: $-90\ V$ for $5\ min$) (b) Comparison of surface potential (black) and surface height (blue) in the region highlighted on the right in (a). (c) The surface topography and (d) local surface potential $U_S$ in the same region as in (b).

3.1 Controlling of charge trapping behaviors

*Controlling the maximum charge density*

Surface charges, in particular controllable surface charges, are favored in many studies, such as energy harvesting[44, 50-51], supercapacitors[52-53], transport of droplets[48], nanofluidics or nanoparticles[54-55], water deionization [56], antifouling[57], and protein adsorption[58]. Since we have now established the charge trapping phenomenon, and have quantitatively characterized the same, it is interesting to explore whether the surface charge density and distributions can be controlled. If so, it would result in a novel, simple and low-cost method of printing controllable surface charges.

To optimize the local surface charge density within the rim, we varied the applied voltage and duration for injecting charges, as well as the polarity of the voltage (Figure 5 and Figure S5). For a fixed charging time of 5 min, the highest charge density of 0.37 mC/m² was achieved at the highest negative charging voltage of -140 V. (The application of higher voltages was hampered by risk of dielectric breakdown. The dielectric strength of the dielectric films is shown in Figure S6). For a safe charging voltage of $-120$ V, the highest charge density as obtained approaching $\sigma_{max} = -0.46\ mC/m^2$ after 20 min. As we have mentioned before, this EW response results show a consistency with the KPFM and local contact angle measurement (labeled in Figures 5a and 5b with highlight circle). Negative charges deposited on the surface were very stable. No appreciable signs of degradation have been observed even after 12 hours of "harsh" testing by continuous probing with a water drop on top of the surface with trapping charges (Figure 5c). After 36 h of immersion in oil, the measured negative charge density was still not altered. The charged AFP films were also placed to a water vapor chamber for 3 h, the obvious difference between the contact angles at the TPCL region and other regions suggested that the trapping charges were still stable in the AFP films. (Figure S7)

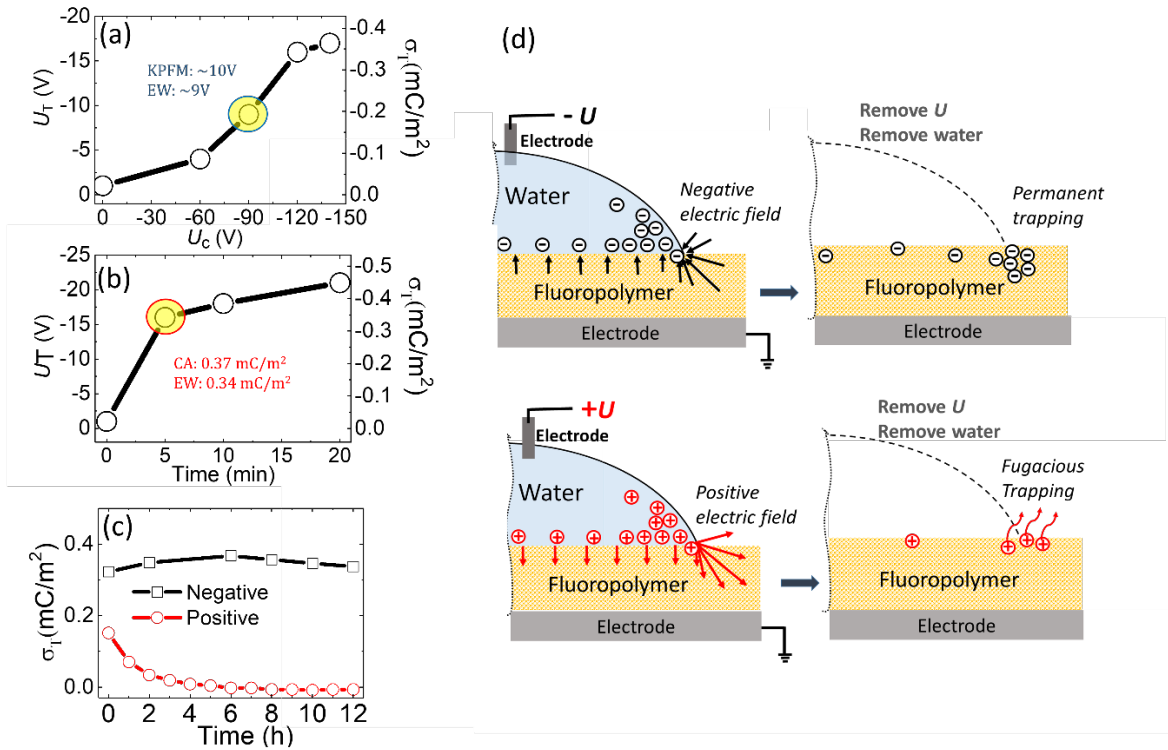

**Figure 5.** Trapping voltage $U_T$ and trapped charge density $\sigma_T$ based on EW measurements as a function of (a) charging voltage ($U_C$) at fixed $t_c = 5\ min$ and (b) charging time $t_c$ at fixed charging voltage ($U_c = -120\ V$). Comparison of $U_T$ and $\sigma_T$ measured by EW, KPFM and local contact angle (CA) of the samples charged at the same conditions are highlighted in (a) and (b). (c) Comparison of the trapping charge density as a function of time (30 V amplitude, 60s period; continuous measurement) between samples charged by -120 V (black) and +120 V (red). The EW curves for (a) and (b), and the corresponding $U_T$ information for (c) are shown in Figure S5. (d) Illustration of charge trapping process at Teflon surface.

In contrast, the maximum positive charge density that we could deposit using the opposite charging polarity was only half of the negative ones, and more importantly, was unstable (Figure 5c). It relaxed within a few hours of continuous probing a water drop in ambient oil. Due to the unstable nature of the positive charges, studies involving these were not explored in more detail. The difference between positive and negative charges suggests a stronger affinity of negative charge carriers to the polymer, which is also indicated in previous results[59-60]. The charge density up to $-0.35 \, mC/m^2$ could be achieved within charging time of $5 \, min$ under a charging voltage of -120V, indicating that 10 times higher charge density than that of the spontaneous charges[37] was reached, and almost hundred times faster than that from spontaneous charge accumulation at Telfon AF surfaces in contact with water at the same pH value[37]. This demonstrates the power of electric fields in immobilizing charge carriers at Teflon-water interfaces. Figure 5d illustrates the charge generation in electrowetting governed by two types of processes: the reversible adsorption and the irreversible trapping. Charges adsorbed in shallow traps on the Teflon surfaces were considered to be reversibly adsorbed. Since Teflon is porous on the molecular scale, traps deeper within the films may also be energetically deeper and hold back ions in an irreversible manner leading to a finite trapped charge. The strong electric fields near the contact lines enable much faster trapping and higher charge densities. In addition, the traps for positive charges are shallower than those for negative ones.

*Creating narrow charge distributions*

According to the classical EW theory,[40-41] the high electric charge densities which should be responsible for the charge injection, are localized within a region of the order of the thickness of the dielectric layer. A solution of the electrostatic problem adapted to the parameters of the present experiments shows that the region, in which the local electric field exceeded the average field $U_c/d$ under the charging drop by more than a factor of two, is less than 1 µm in width, as shown in Figure 6. Following this thought, the intrinsic charge generation mechanism should allow to generate much narrower charge distribution region than the measured width of 200 to 300 µm. We attribute this to the relaxation of the contact line position during the charging process that accompanies the contact angle relaxation (Figures 1b and c).

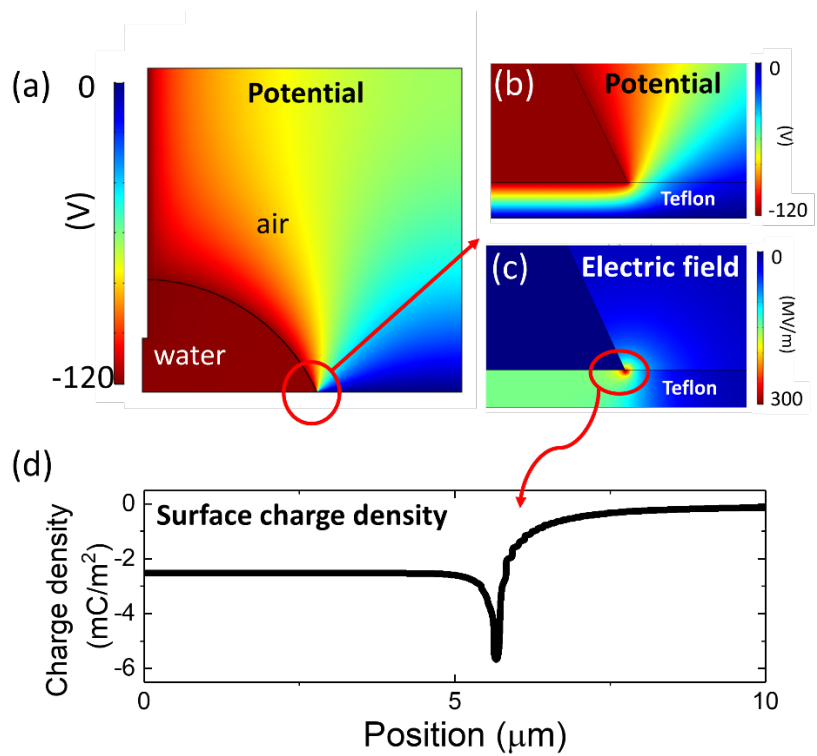

**Figure 6.** Fringe effect simulation using finite element method (on Comsol platform)(a) of the potential of the whole simulation region, (b) the potential, (c) the electric field and (d) the charge density on the drop near the contact line. For details of the simulations, see S.I. and Figure S8.

The slowly receding contact line was believed to leave behind a trace of charges on the surface, which eventually formed the observed rim. In order to reduce the width of the deposited rim of charges, we suppressed the geometric relaxation of the drop during charging by confining it between two parallel plates at a distance of $h = 100$ μm (Figure 7a and 7b). $h$ was simply achieved by putting a 100 μm thick glass space between the two plates. The lower surface was a Teflon-coated substrate as above and the upper one was an ITO coated glass that served as an electrode during the charging process. These two surfaces confined a drop during charging and reduce the displacement of the contact line $\Delta R$ for the same amount of contact angle relaxation $\Delta \theta$ as in Figure 1, with $\Delta R \propto h\, \Delta(\cos \theta)$. KPFM measurements after removing the top surface and the drop demonstrate that indeed a much narrower rim of charges was deposited with a width of about 20 μm, as shown in Figure 7. The average surface potential within the rim was -10 V, corresponding to the trapped charge density $\sigma_T = -0.22\ mC/m^2$. From these results, we could also conclude that a much smaller but finite charge density was deposited at the solid-liquid interface away from the contact line, as seen in Figure 4. However, the edge of the charged rim was still sharp, suggesting that a further reduction of the width of the rim towards the intrinsic limit should be possible. Thus, by altering the charging voltage, charging time and manipulating the TPCL motions, this EW-induced charge trapping phenomenon was demonstrated as a promising strategy to fabricate surface charges with various density at microscale.

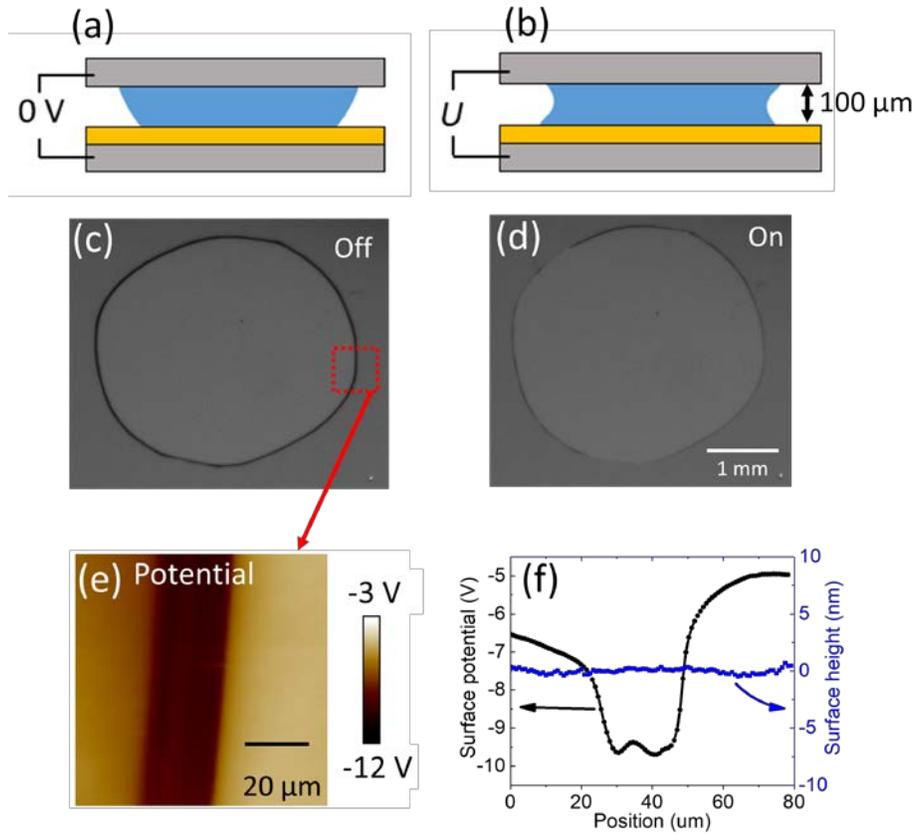

**Figure 7.** Generation of narrow charge patterns using a confined drop with $U_C = -90\ V$ for 5 min. (a) and (b) schematic setup in voltage off and on state. (c) and (d) corresponding drop views in transmission illustrating the small variation in radius. (e) KPFM surface potential map of the region indicated in (c). (f) line profiles of surface topography and surface potential of charged region in (e).

## 4. CONCLUSION AND OUTLOOK

In this work, we reveal that, the deposition of surface charges from an aqueous drop on an electrically insulating fluoropolymer surface in an EW configuration at high voltage, indeed preferentially occurred along the contact line of the drop, in accordance with the established EW theory but deviating from recent suggestions based on unconventional and/or unstable EW systems[32, 42-43]. This observation was confirmed by a combination of three experimental techniques, namely local contact angle at zero voltage, the asymmetry of the EW response, and KPFM. We thus proposed this EW-induced charge trapping method as a simple and low-cost strategy for generating surface charges at microscale. We have demonstrated the tunability of the charge density and the charged regions by simply adjusting the electric charging conditions or manipulating the motion of TPCL. No vacuum process or other complex facilities were needed. Typically, the positive charges temporally stayed at fluoropolymer surfaces; however, the negative charges showed extremely high stability for long time even in water or humid environment. We believe this novel surface charge depositing strategy will be beneficial to a wide range of research and applications which require controllable surface charges.

ASSOCIATED CONTENT

**Supporting Information**.

Illustration of charging process in humid chamber. EW response curve of Teflon AF film with 800 nm thickness. Measuring trapping voltage by KPFM. Setup of local contact angle measurement. EW response curve of various charging conditions. Leakage current measurement. Trapping charge testing after water vapor treatment for 3h. Details of simulations. (PDF)

Video of charging process under -120 V for 5 min at 30 times speed. (AVI)

Video of local contact angel testing at 8 times speed. (AVI)

AUTHOR INFORMATION

**Corresponding Authors**

*Email: hao.wu@utwente.nl

*E-mail: f.mugele@utwente.nl

**Present Addresses**

[e] Dynamics of Complex Fluids, Max Planck Institute for Dynamics and Self-organization, Am Fassberg 17, 37077 Goettingen, Germany.

ACKNOWLEDGMENT

This work was supported by National Key R&D Program of China (2016YFB0401501), National Natural Science Foundation of China (Grant No. U1501244, U1601651), Program for Chang Jiang Scholars and Innovative Research Teams in Universities (No. IRT_17R40), Science and Technology Project of Shenzhen Municipal Science and Technology Innovation Committee (GQYCZZ20150721150406), Guangdong Provincial Key Laboratory of Optical Information Materials and Technology (No. 2017B030301007), MOE International Laboratory for Optical Information Technologies and the 111 Project.

**Supporting information:**

# Electrically controlled localized charge trapping at amorphous fluoropolymer-electrolyte interfaces

*Hao Wu* [*,a,b,c], *Ranabir Dey* [c,e], *Igor Siretanu* [c], *Dirk van den Ende* [c], *Lingling Shui* [a, b], *Guofu Zhou* [a,b,d], *Frieder Mugele* [*,c]

[a] Guangdong Provincial Key Laboratory of Optical Information Materials and Technology & Institute of Electronic Paper Displays, South China Academy of Advanced Optoelectronics, South China Normal University, Guangzhou 510006, P. R. China.

[b] National Center for International Research on Green Optoelectronics, South China Normal University, Guangzhou 510006, P. R. China.

[c] Physics of Complex Fluids, Faculty of Science and Technology, MESA+ Institute for Nanotechnology, University of Twente, Enschede 7500AE, The Netherlands

[d] Shenzhen Guohua Optoelectronics Tech. Co. Ltd., Shenzhen 518110, P. R. China

[e] Present address: Dynamics of Complex Fluids, Max Planck Institute for Dynamics and Self-organization, Am Fassberg 17, 37077 Goettingen, Germany.

1. **Charging Teflon film in humid environment**

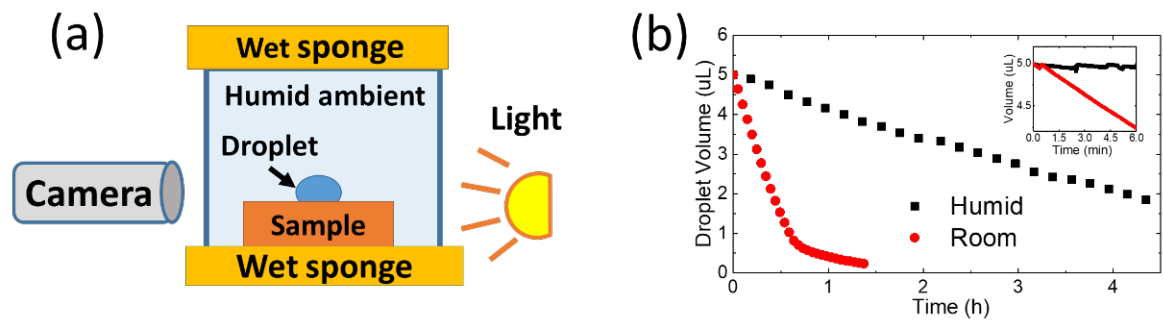

**Figure S1**. (a) Schematic drawing of humid chamber setup. (b) comparison of water droplet evaporation rate in room environment and in a humid chamber.

2. **Electrowetting response**

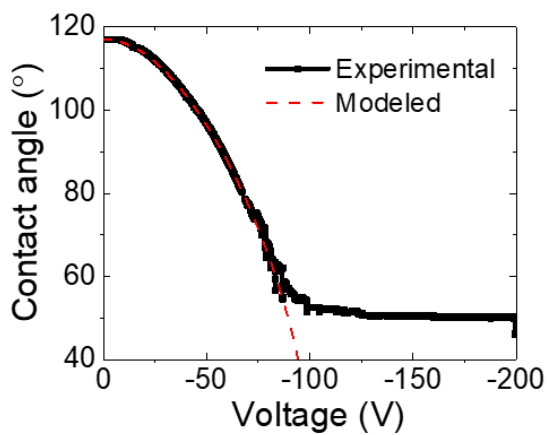

**Figure S2.** Contact angle depending on applied voltage. Modeled curve is calculated with Young-Lippmann equation.

## 3. Measuring trapping voltage ($U_T$) by Kelvin probe force microscopy (KPFM)

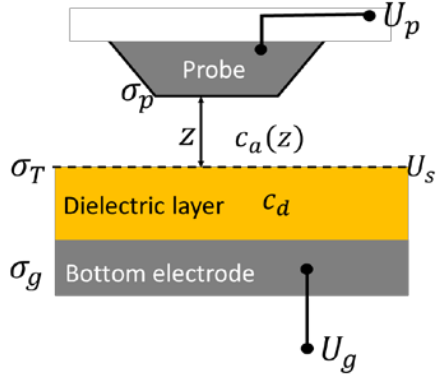

**Figure S3.** The schematic of measuring the trapping voltage ($U_T$) and trapping charge density $\sigma_T$ by utilizing KPFM. A detailed explanation is given below.

According to Fig. S3 the charge density on the probe of KPFM ($\sigma_p$) and on the bottom electrode ($\sigma_g$) are

$$\sigma_p = c_a(z)(U_p - U_s) \quad (S1)$$

$$\sigma_g = c_d(U_g - U_s) \quad (S2)$$

where the $c_a(z)$ is the capacitance per unit area of the air capacitor formed between the probe and the dielectric surface. $c_a(z)$ depends on the distance between the probe and the dielectric layer (z), and can be calculated by $c_a(z) = \varepsilon_0/z$. $c_d$ is the capacitance per unit area of the dielectric layer. $U_p$, $U_s$ and $U_g$ are the potential at the probe, at the dielectric surface and the bottom electrode. Given the total charge in the system is zero, we get

$$\sigma_g + \sigma_p + \sigma_T = 0 \quad (S3)$$

where $\sigma_T$ is the trapping charge density on the surface of dielectric layer.

From Eq. S1 – S3, we get:

$$U_s - U_p = \frac{\sigma_T}{c_a(z) + c_d} - \frac{c_d}{c + c_d}(U_p - U_g) \tag{S4}$$

$$U_s - U_g = \frac{\sigma_T}{c_a(z) + c_d} + \frac{c_a}{c + c_d}(U_p - U_g) \tag{S5}$$

Here, we should notice that the surface potential of the dielectric layer

$$U_s = \frac{1}{c_a(z)+c_d}\left(\sigma_T + c_a(z)U_p + c_d U_g\right) \tag{S6}$$

is not constant during the measurement. The dielectric surface potential $U_s(z, U_p)$ does not only depend on the trapping charge density $\sigma_T$, but also on z and $U_p$. We define the trapping voltage as $U_T = \sigma_T/c_d$. Only if the probe is very far away from the dielectric surface, the surface potential equals the trapping voltage, $U_s = U_T = \sigma_T/c_d$. The electric energy of the system $W_{el}$ contain two parts: the energy from the capacitance $W_c$ and the energy from the source (battery) $W_s$, can be written as:

$$W_{el} = W_c + W_s \tag{S7}$$

The energy of the capacitance contains the energy in the air capacitance between the probe and the surface ($W_a$) and the energy in the dielectric layer $W_d$, and can be calculated as:

$$W_c = W_a + W_d = \frac{1}{2}Ac_a(z)(U_p - U_s)^2 + \frac{1}{2}Ac_d(U_s - U_g)^2$$

$$= \frac{1}{2}A\left[\frac{\sigma_T^2}{c_a(z) + c_d} + \frac{c_a(z)c_d}{c_a(z) + c_d}(U_p - U_g)^2\right] \tag{S8}$$

The energy from the source is:

$$W_s = W_0 - \sigma_p A(U_p - U_g) \tag{S9}$$

Where $W_0$ is the initial energy stored in the source. $A$ is the overlapping area of the probe and the dielectric surface. According to Eq. S1, S2 and S3, the charge density on the probe is:

$$\sigma_p = c_a(z)\left(\frac{-\sigma_T}{c_a(z) + c_d} + \frac{c_d}{c_a(z) + c_d}(U_p - U_g)\right) \tag{S10}$$

Substitute Eq. S10 to Eq. S9, we get

$$W_s = W_0 - Ac_a\left[\frac{-\sigma_T}{c_a + c_d} + \frac{c_d}{c_a + c_d}(U_p - U_g)\right](U_p - U_g)$$

$$= W_0 + A\frac{c_a \sigma_T}{c_a + c_d}(U_p - U_g) - A\frac{c_a c_d}{c_a + c_d}(U_p - U_g)^2 \tag{S11}$$

Substitute Eq. S8 and Eq. S11 to Eq. S7, the electric energy in the system is

$$W_{el} = \frac{1}{2}A\left[\frac{\sigma_T^2}{c_a + c_d} + \frac{c_a c_d}{c_a + c_d}(U_p - U_g)^2\right] + W_0 + A\frac{c_a \sigma_T}{c_a + c_d}(U_p - U_g) - \frac{c_a c_d}{c_a + c_d}(U_p - U_g)^2$$

$$= W_0 + \frac{1}{2}A\left(\frac{\sigma_T^2}{c_a + c_d}\right) - \frac{1}{2}A\frac{c_a c_d}{c_a + c_d}(U_p - U_g)^2 + A\frac{c_a \sigma_T}{c_a + c_d}(U_p - U_g) \tag{S12}$$

Given the trapping voltage $U_T = \sigma_T/c_d$ and the total capacitance $C(z)$ of the probe and the dielectric layer:

$$C(z) = A \cdot \frac{c_a(z) c_d}{c_a(z) + c_d} \tag{S13}$$

the electrical energy is given by:

$$W_{el} = W_0 - \frac{1}{2}C(z)(U_p - U_g - U_T)^2 + \frac{1}{2}Ac_dU_T^2 \qquad (S14)$$

The electric force on the AFM probe is given by the gradient of the energy:

$$F_{el} = -\frac{\partial W_{el}}{\partial z} = \frac{1}{2}\frac{\partial C(z)}{\partial z}(U_p - U_g - U_T)^2 \qquad (S15)$$

Because $\frac{\partial C}{\partial z}$ is negative, this force is attractive. Since the potential on the tip is the sum of an AC voltage ($U_{AC}\sin\omega t$) and a DC voltage ($U_{DC}$), $U_p$ can be written as:

$$U_p = U_{DC} + U_{AC}\sin\omega t \qquad (S16)$$

while the bottom electrode is grounded: $U_g = 0$. Thus, the electric force is:

$$F_{el} = \frac{1}{2}\frac{\partial C(z)}{\partial z}(U_{DC} - U_T + U_{AC}\sin\omega t)^2 \qquad (S17)$$

$$= \frac{1}{2}\frac{\partial C(z)}{\partial z}\left[(U_{DC} - U_T)^2 + 2(U_{DC} - U_T)U_{AC}\sin\omega t + \frac{1}{2}U_{AC}(1 - \cos 2\omega t)\right]$$

Splitting the force according to its frequency components, we obtain the static ($F_{DC}$) and dynamic ($F_\omega$ and $F_{2\omega}$) contributions:

$$F_{DC} = \frac{\partial C(z)}{\partial z}\left[\frac{1}{2}(U_{DC} - U_T)^2 + \frac{1}{4}U_{AC}^2\right] \qquad (S18)$$

$$F_\omega = \frac{\partial C(z)}{\partial z}(U_{DC} - U_T)U_{AC}\sin\omega t \qquad (S19)$$

$$F_{2\omega} = -\frac{1}{4}\frac{\partial C(z)}{\partial z}U_{AC}\cos 2\omega t \qquad (S20$$

4. **Schematic drawing of local contact angle measurement by probe-droplet**

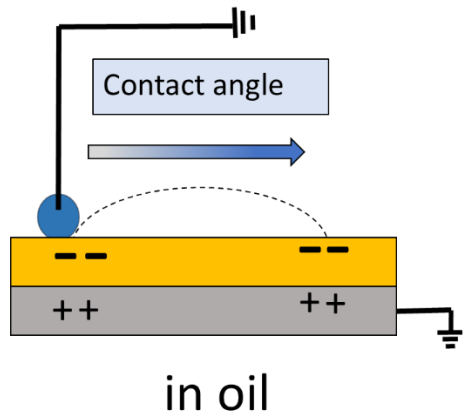

**Figure S4**. Schematic drawing of local contact angle measurement by probe-droplet.

## 5. Charge trapping contorted by charging voltage and time

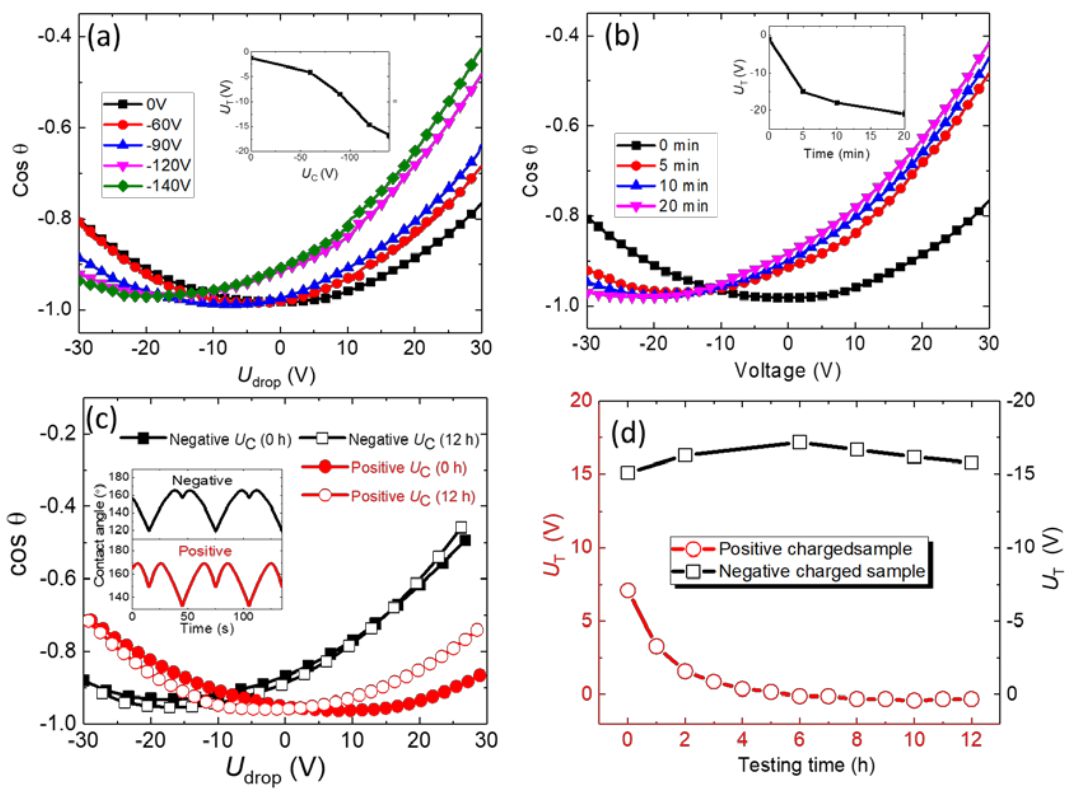

**Figure S5.** The electrowetting curve of samples charged at (a) various voltage for 5 min and (b) -120 V for various time. Inserts in (a) and (b) are the trapping voltage dependence of the charging voltage and charging time. (c) Electrowetting response comparison of two charged samples under -120 and +120 V, respectively, for 5 min. (d) Trapping voltage varying with time driven by a triangular waveform (30 V amplitude and 60 s period).

## 6. Leakage current measurement

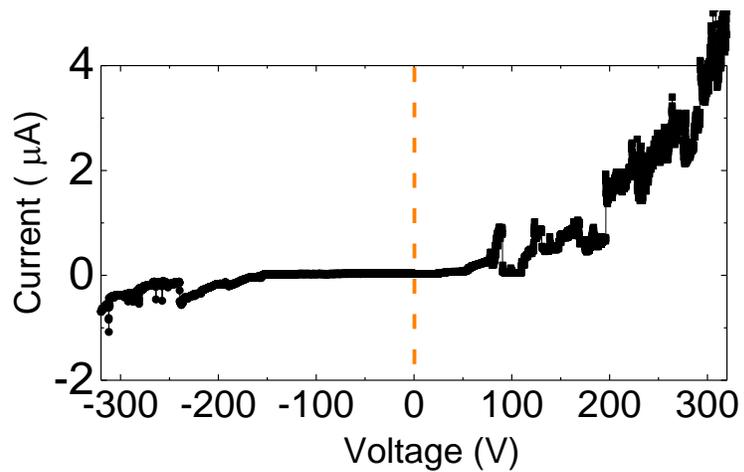

**Figure S6**. Leakage current depending on applied voltage of sample with a 0.8 µm thick Teflon films.

## 7. Charge trapping detected after being treated with water vapor

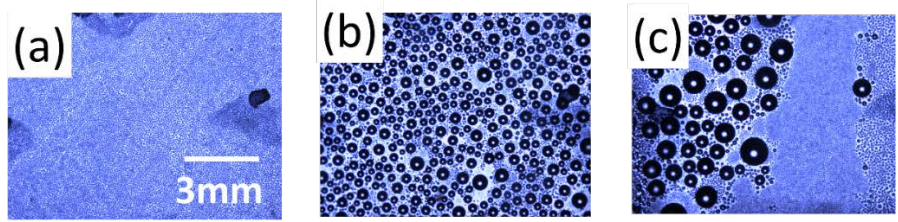

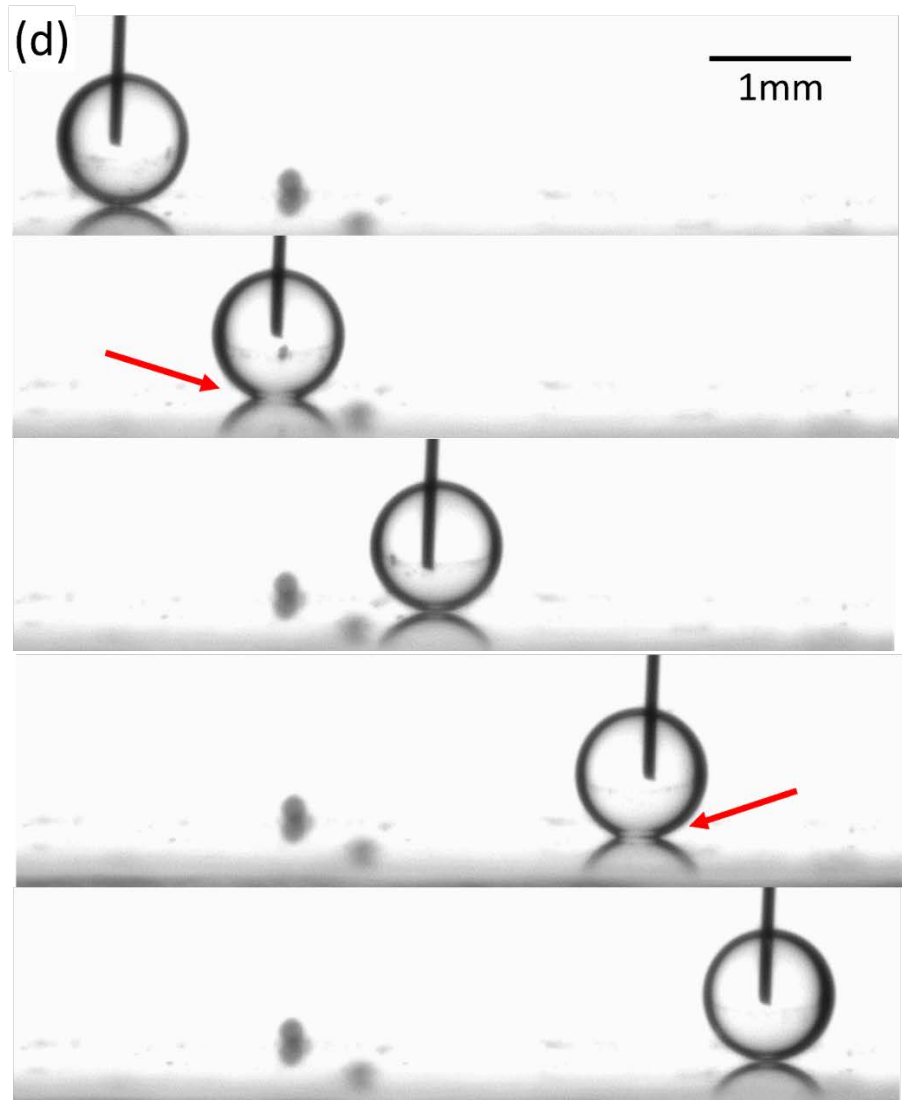

**Figure S7**. (a-c) The image of condensation phenomenon during charge sample (charged by -120 V for 5 min) being treated with water vapor for 3 h. The sample is vertical placed in a humid chamber with water vapor. The water vapor forms (a) small droplets, then (b) the droplets grow bigger, and (c) slide down and now droplet forms on the sample surfaces. (d) Snapshots of local contact angle merriment at different position on the surface of the sample treated by water vapor for 3 h. The red arrows point the TPL region with trapping charges.

## 8. Information of fringe effect simulation

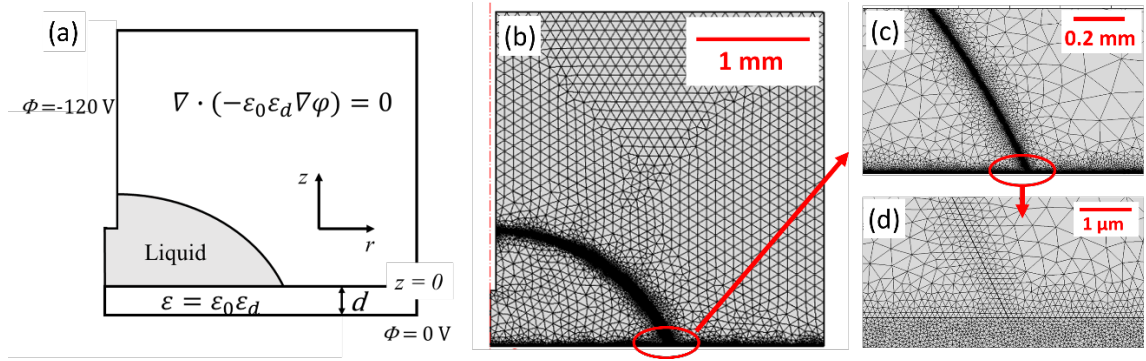

**Figure S8**. (a) Schematic showing the axi-symmetric geometry used for the numerical simulations (b-d) Meshes used for the finite element based simulation using the Comsol platform. The simulation box is 3 mm × 3 mm in size, and the axis of symmetry is represented by the dash-dotted line in (b). The thickness of the dielectric layer is 800 nm. The size of the elements for the dielectric layer is 10 nm to 100 nm, and both for the droplet and air domains it is 10 nm to 100 µm. The droplet is considered perfectly conductive. The two sides of the platinum electrode, dipped into the droplet, are considered to be at -120 V (Dirichlet boundary condition), and the electrode underneath the dielectric is considered to be at 0 V (Dirichelt boundary condition). The outer domain boundaries are maintained at the no flux condition (Neumann boundary condition). The potential and the field distributions are calculated by solving the equation $\nabla \cdot (\varepsilon_0 \varepsilon_r \nabla \varphi) = 0$ in each of the domains (dielectric, droplet and air) where $\varepsilon_0$ is the vacuum permittivity; $\varepsilon_r$ is the relative permittivity of the insulator (1.93 for Teflon in this case), and $\varphi$ is the electric potential.